\documentclass[10pt]{article}
\usepackage{graphicx} 

\newcommand{\ba}{\begin{eqnarray}}
\newcommand{\ea}{\end{eqnarray}}

\begin{document}

\parindent=1.0cm 
\begin{center}

{\large \bf Quantum Phase Transitions in a Finite System}

\bigskip
{\large A. Leviatan}

{Racah Institute of Physics, The Hebrew University,
Jerusalem 91904, Israel}

\end{center}

\centerline{Abstract}
{\it A general procedure for studying finite-N effects 
in quantum phase transitions of finite systems is presented and
applied to the critical-point dynamics of nuclei 
undergoing a shape-phase transition of 
second-order (continuous), and of first-order 
with an arbitrary barrier.}

\bigskip

An important issue concerning quantum phase transitions in mesoscopic 
systems is to understand the modifications at criticality due to 
their finite number of constituents. In the present contribution we 
study this question in connection with nuclei exemplifying a finite 
system undergoing a shape-phase transition. We employ the 
interacting boson model (IBM)~\cite{ibm} which describes low-lying 
quadrupole collective states in nuclei in terms of a system of $N$ 
monopole ($s$) and quadrupole ($d$) bosons representing valence nucleon pairs. 
The model is based on a $U(6)$ spectrum generating algebra and its three 
dynamical symmetry limits: $U(5)$, $SU(3)$, and $O(6)$, 
describe the dynamics of stable nuclear shapes: spherical, axially-deformed, 
and $\gamma$-unstable deformed.
A geometric visualization of the model is obtained by 
an intrinsic energy surface defined by 
the expectation value of the Hamiltonian in the coherent (intrinsic) 
state~\cite{gino80,diep80}
\ba
\vert\,\beta,\gamma ; N \rangle &=&
(N!)^{-1/2}(b^{\dagger}_{c})^N\,\vert 0\,\rangle ~,
\label{cond}
\ea
where $b^{\dagger}_{c} = (1+\beta^2)^{-1/2}[\beta\cos\gamma\,
d^{\dagger}_{0} + \beta\sin{\gamma}\,
( d^{\dagger}_{2} + d^{\dagger}_{-2})/\sqrt{2} + s^{\dagger}\,]$.
For the general IBM Hamiltonian with one- and two-body interactions, the 
energy surface takes the form
\ba
E(\beta,\gamma) &=& E_0 + 
N(N-1)(1+\beta^2)^{-2}\,
\left [ a\beta^{2} - b\beta^3\cos 3\gamma + c\beta^4\right ]~.
\label{eint}
\ea
The coefficients $E_0,a,b,c$ involve particular linear 
combinations of the Hamiltonian's parameters~\cite{lev87}. 
The quadrupole shape parameters in the 
intrinsic state characterize the associated equilibrium shape. 
Phase transitions for finite N can be studied 
by an IBM Hamiltonian involving terms from different dynamical 
symmetry chains~\cite{diep80}. 
Several works have followed this route in numerical studies of finite-N 
effects at criticality [5-8]. 
In the present contribution we consider an 
(approximate) analytic-oriented approach to this problem [9-11].  

The nature of the phase transition is 
governed by the topology of the corresponding intrinsic 
energy surface which serves as a Landau's potential. In a 
second-order phase transition, the energy surface is $\gamma$-independent 
and has a single minimum which changes continuously from a spherical to 
a deformed $\gamma$-unstable phase. At the critical-point $a=b=0$ and 
the energy-surface acquires a flat behaviour ($\sim \beta^4$) for small 
$\beta$ (justifying the use of a square-well potential in the 
E(5) critical-point model~\cite{iac00}). 
This is the situation encountered 
in the $U(5)$-$O(6)$ phase transition 
where the critical Hamiltonian is given by 
\ba
H_{cri} &=& \epsilon\,\hat{n}_d + {1\over 4}A\,
\Bigl [\, d^{\dagger}\cdot d^{\dagger} -  (s^{\dagger})^2\,\Bigr ]
\left[\, \tilde{d}\cdot \tilde{d} -  s^2\,\right ]
\quad , 
\quad 
\epsilon = (N-1)A ~.
\label{hcri2nd}
\ea
Here $\tilde{d}_{\mu}=(-1)^{\mu}d_{-\mu}$ 
and the dot implies a scalar product. 
$H_{cri}$ is $O(5)$-invariant and involves the $U(5)$ term $\hat{n}_d$ (the
$d$-boson number operator), and the $O(6)$-pairing term. 
The intrinsic energy surface of $H_{cri}$ has the form
\ba
E(\beta) &=& \frac{1}{4}AN(N-1) + A\,N(N-1)\,\beta^4(1+\beta^2)^{-2} ~.
\label{enesurf2}
\ea
As shown in Fig.~(1a), 
The global minimum at $\beta=0$ is not well-localized
and $E(\beta)$ exhibits considerable instability in $\beta$.
Under such circumstances fluctuations in $\beta$ are large and
play a significant role in the dynamics. Some of their effect can
by can be taken into account by means of variation after projection of 
states of good $O(5)\supset O(3)$ symmetry $\tau LM$ from the coherent 
state in Eq.~(\ref{cond})
\ba
\vert\, \xi=1;\beta,N,\tau, L,M\rangle &\propto& 
\left [F_{N}^{(\tau)}(\beta)\right ]^{-1/2} 
\hat{\cal{P}}_{\tau,LM}\vert \beta,\gamma; N\rangle
\nonumber\\
F_{N}^{(\tau)}(\beta) &=&  
\sum_{n_d=\tau}^{N}{1\over 2}\left [1 + (-1)^{n_d-\tau}\right ]
\frac{\beta^{2n_d}}{(N-n_d)!\,(n_d-\tau)!!\,(n_d+\tau+3)!!} ~.
\label{wfqpt2}
\ea
These $\tau$-projected states form the ground band ($\xi=1$) and 
interpolate between the $U(5)$ ground state, 
$\vert\,N,n_d=\tau=L=0\rangle\equiv \vert s^N\rangle$ at $\beta=0$ 
and the $O(6)$ 
ground band, $\vert\,N,\sigma=N,\tau,L\rangle$ at $\beta=1$.
\begin{figure}[pt]  
\centerline{
\rotatebox{270}{\includegraphics[height=90mm]{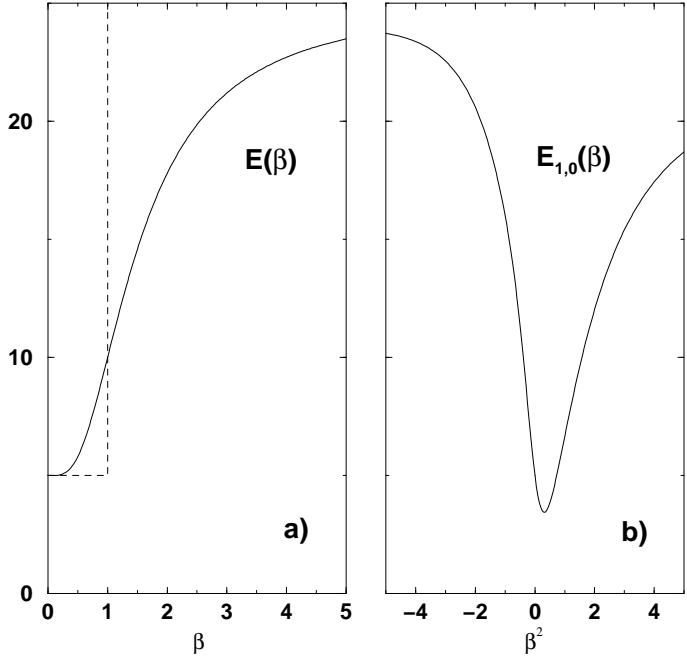}}}
\caption{Energy surfaces of the critical IBM Hamiltonian $H_{cri}$
(\ref{hcri2nd}) with $N=5$ and $A=1$.
(a)~Intrinsic energy surface $E(\beta)$, Eq.~(\ref{enesurf2})
[solid line], and its approximation by a
square-well potential [dashed line].
(b)~$O(5)$ projected energy surface $E_{\xi=1,\tau=0}(\beta)$,
Eq.~(\ref{enetau}) whose global minimum is at $\beta^2=0.314$.
\label{f2}}
\end{figure}
\begin{table}[b]
\centering
\caption{
Excitation energies (in units of $E(2^{+}_{1,1}=1$) 
and B(E2) values (in units of
$B(E2;\, 2^{+}_{1,1}\to 0^{+}_{1,0})=1$) 
for the E(5) model {\protect\cite{iac00}}, 
for several N=5 calculations and for the experimental data of 
$^{134}$Ba. 
The finite-N calculations involve 
the exact diagonalization of the critical 
IBM Hamiltonian ($H_{cri}$) [Eq.~(\ref{hcri2nd})], 
$\tau$-projected states, $L^{+}_{\xi,\tau}$, 
[Eq.~(\ref{wfqpt2}) with $\beta^2=0.314$], 
$U(5)$ and $O(6)$ limits of the IBM. 
\normalsize}
\medskip
\begin{tabular}{lcccccc}
\hline\hline\noalign{\smallskip}
& E(5) & exact &
$\tau$-projection & $U(5)$ & $O(6)$ & $^{134}$Ba \\
&  & N=5 & N=5 & N=5  & N=5  & exp \\
\noalign{\smallskip}\hline\noalign{\smallskip}
$E(L^{+}_{1,2})$ & 2.20 & 2.195 & 2.19  & 2  & 2.5 & 2.32 \\
$E(L^{+}_{1,3})$ & 3.59 & 3.55  & 3.535 & 3  & 4.5 & 3.66 \\
$E(0^{+}_{2,0})$ & 3.03 & 3.68  & 3.71  & 2  & 
$1.5{A\over B}$ & 3.57  \\
\noalign{\smallskip}\hline\noalign{\smallskip}
$B(E2;\, 4^{+}_{1,2}\to 2^{+}_{1,1})$ &
1.68 & 1.38 & 1.35 & 1.6 & 1.27 & 1.56(18) \\
$B(E2;\, 6^{+}_{1,3}\to 4^{+}_{1,2})$  & 
2.21 & 1.40 & 1.38 & 1.8 & 1.22 &  \\
$B(E2;\, 0^{+}_{2,0}\to 2^{+}_{1,1})$ & 
0.86 & 0.51 & 0.43 & 1.6 & 0 & 0.42(12) \\
\noalign{\smallskip}\hline\hline
\end{tabular}
\end{table}
The matrix element of 
$H_{cri}$~(\ref{hcri2nd}) in these 
states defines the $\tau$-projected 
energy surface which can be evaluated in closed form~\cite{levgin03}
\ba
E^{(N)}_{\xi=1,\tau}(\beta) &=&
\epsilon\left[\, N - S^{(N)}_{1,\tau}\,\right ]
+ {1\over 4}A\, (1-\beta^2)^2\, S^{(N)}_{2,\tau} ~.
\label{enetau}
\ea
Here $S^{(N)}_{k,\tau}$ denote the expectation values of
$(s^{\dagger})^ks^k$ in the states (\ref{wfqpt2}), and are given by
$S^{(N)}_{1,\tau} = F^{(\tau)}_{N-1}/F^{(\tau)}_{N}$
and $S^{(N)}_{2,\tau} = S^{(N)}_{1,\tau}S^{(N-1)}_{1,\tau}$.
Members of the first excited band $(\xi=2)$ have approximate
wave functions of the form
\ba
\vert\, \xi=2; \beta, N,\tau, L\rangle &=& 
{\cal N}_{\beta}\,P^{\dagger}_{\beta}\,\vert\, \xi=1; \beta,N-2,\tau,L\rangle
\ea
with $P^{\dagger}_{\beta}= 
[d^{\dagger}\cdot d^{\dagger} - \beta^2(s^{\dagger})^2]$ 
and ${\cal N}_{\beta}$ a known normalization. 
Explicit expressions for quadrupole rates involving $\tau$-projected states 
can be derived. 
For example, with the 
$E2$ operator
$T(E2) = d^{\dagger}s + s^{\dagger}\tilde{d}$, 
\ba
&&B(E2;\, \xi=1;\, \tau+1, L^{\prime}=2\tau+2
\to \xi=1,\, \tau,L=2\tau) =
\frac{(\tau+1)}{(2\tau+5)}
\frac{\beta^2 \left [F_{N-1}^{(\tau)}(\beta) 
+ F_{N-1}^{(\tau+1)}(\beta)\right]^2}
{F_{N}^{(\tau)}(\beta)\,F_{N}^{(\tau+1)}(\beta)} ~.
\ea
Similar expressions for the excited-band energies
$E^{(N)}_{\xi=2,\tau}(\beta)$ and interband $E2$ rates are 
available~\cite{levgin03}. 
As seen from Fig.~(1), in contrast to $E(\beta)$, 
the lowest $O(5)$ projected energy surface $E_{\xi=1,\tau=0}(\beta)$ 
supports a well-defined 
global minimum at a certain value of $\beta$. 
As shown in Table~1, using this effective $\beta$-deformation in the 
$\tau$-projected states provides 
\begin{figure}[pt]  
\centering
\rotatebox{270}{\includegraphics[width=0.4\linewidth]{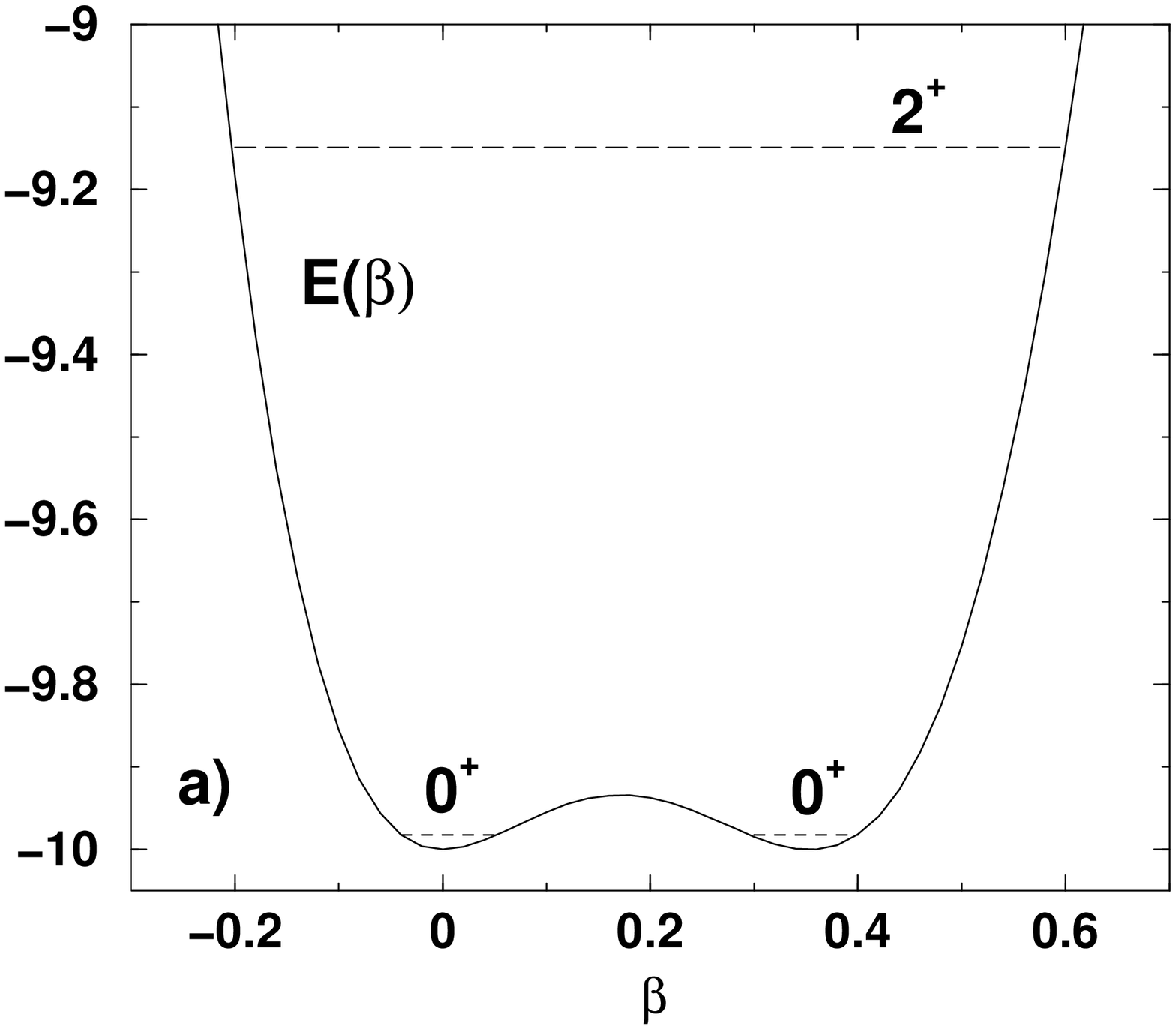}}
\rotatebox{270}{\includegraphics[width=0.4\linewidth]{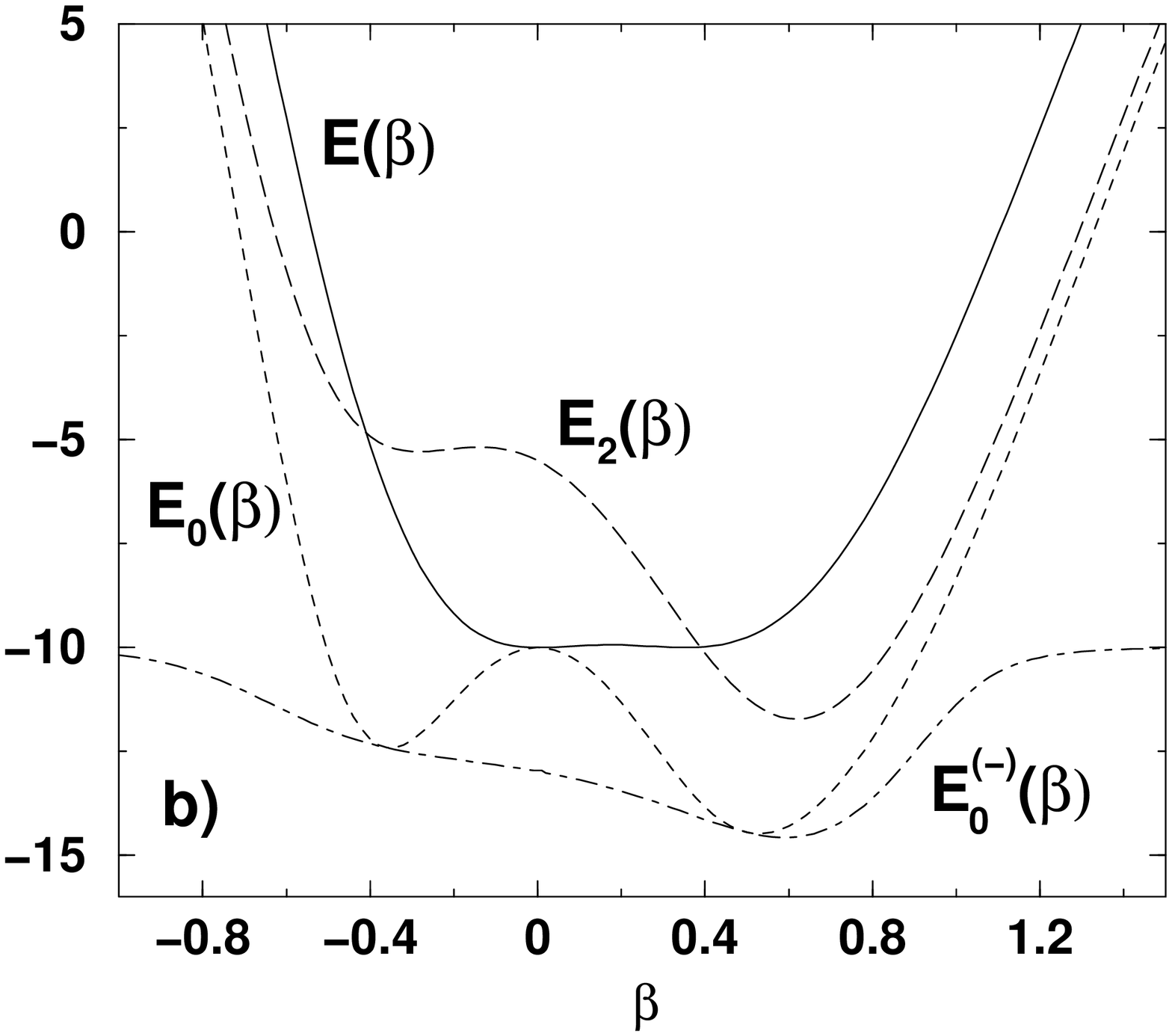}}
\caption{Energy surfaces
of the critical IBM Hamiltonian, Eq.~(\ref{hcri1st}), 
with $\kappa=0.2$ and $N=10$. 
(a)~Intrinsic energy surface
$E(\beta)\equiv E(\beta,\gamma=0)$,
Eq.~(\ref{enesurf}), [solid line]. 
The unmixed $L=0$ and $L=2$ levels are shown for illustration. 
(b) $E(\beta)$ [solid line] as in~(a), 
unmixed $L=0$ [dashed line] and $L=2$ 
[long dashed line] projected energy surfaces, 
$E_{L}(\beta)\equiv E^{(N)}_{L}(\beta)$, 
Eq.~(\ref{egL}), and 
the lowest $L=0$ eigenpotential [dot-dashed line], 
$E^{(-)}_{L=0}(\beta)$, whose global minimum is at 
$\beta=0.591$.
\label{f3}}
\end{figure}
\begin{table}[bt]
\centering
\caption{
Excitation energies (in units of $E(2^{+}_{1})=1$) 
and B(E2) values (in units of
$B(E2;\, 2^{+}_{1}\to 0^{+}_{1})=1$) 
for the X(5) critical-point model {\protect\cite{iac01}}, 
for several N=10 calculations, and for the experimental data of 
$^{152}$Sm. The finite-N calculations involve 
the exact diagonalization of the critical 
IBM Hamiltonian [Eq.~(\ref{hcri1st})], $L$-projected states 
[Eqs.~(\ref{wfqpt1}) with $\beta= 0.591$], 
$U(5)$ and $SU(3)$ limits of the IBM. 
\normalsize}
\vskip 10pt
\begin{tabular}{lcccccc}
\hline\hline\noalign{\smallskip}
& X(5) & exact & $L$-projection & $U(5)$ & $SU(3)$ & $^{152}$Sm \\
&      & N=10  & N=10           & N=10   & N=10   & exp \\
\noalign{\smallskip}\hline\noalign{\smallskip}
$E(4^{+}_{1})$  & 2.91  & 2.43 & 2.46 & 2  & 3.33  & 3.01 \\
$E(6^{+}_{1})$  & 5.45  & 4.29 & 4.33 & 3  & 7.00  & 5.80 \\
$E(8^{+}_{1})$  & 8.51  & 6.53 & 6.56 & 4  & 12.00 & 9.24  \\
$E(10^{+}_{1})$ & 12.07 & 9.12 & 9.13 & 5  & 18.33 & 13.21 \\
$E(0^{+}_{2})$  & 5.67  & 2.64 & 3.30 & 2  & 25.33 & 5.62   \\
\noalign{\smallskip}\hline\noalign{\smallskip}
$B(E2;\, 4^{+}_{1}\to 2^{+}_{1})$ &
1.58 & 1.61 & 1.60 & 1.8 & 1.40 & 1.45 \\
$B(E2; \, 6^{+}_{1}\to 4^{+}_{1})$ & 
1.98 & 1.85 & 1.80 & 2.4 & 1.48 & 1.70  \\
$B(E2;\, 8^{+}_{1}\to 6^{+}_{1})$ & 
2.27 & 1.92 & 1.87 & 2.8 & 1.45 & 1.98 \\
$B(E2;\, 10^{+}_{1}\to 8^{+}_{1})$  & 
2.61 & 1.87 & 1.86 & 3.0 & 1.37 & 2.22  \\
$B(E2;\, 0^{+}_{2}\to 2^{+}_{1})$ & 
0.63 & 0.78 & 0.61 & 1.8 & 0.07 & 0.23 \\
\noalign{\smallskip}\hline\hline
\end{tabular}
\end{table}
accurate analytic estimates
to the exact finite-N calculations of the critical
IBM Hamiltonian which in-turn 
capture the essential features of the E(5) critical-point 
model~\cite{iac00} present in $^{134}$Ba. 

In a first-order phase transition the intrinsic energy surface has two 
coexisting minima which become degenerate at the critical-point. 
This is the case 
in the 
$U(5)$-$SU(3)$ 
phase 
transition 
where the critical 
Hamiltonian
\ba
H_{cri} &=& \epsilon\,\hat{n}_d -\kappa\, Q\cdot Q
\quad ,\quad
\epsilon = \frac{9}{4}\kappa (2N-3)
\label{hcri1st}
\ea
involves the $U(5)$ pairing and the $SU(3)$ quadrupole terms. 
The associated 
intrinsic energy surface
\ba
E(\beta,\gamma) &=& 
-5\kappa N +
\frac{\kappa\,N(N-1)\,\beta^2}{2(1+\beta^2)^{2}}
\left ( 1 -4\sqrt{2}\beta\cos3\gamma + 8\beta^2 \right ), 
\label{enesurf}
\ea 
has two degenerate minima, at $\beta=0$ and at 
$(\beta=\frac{1}{2\sqrt{2}},\gamma=0)$. 
As shown in Fig.~(2),   
the barrier separating the spherical and prolate-deformed minima is 
extremely small and the resulting surface, 
$E(\beta)\equiv E(\beta,\gamma=0)$, is rather flat. 
This behaviour motivated the use of a square-well potential in the 
X(5) model~\cite{iac01}. 
Particularly relevant are states of 
good $O(3)$ symmetry $L$ 
projected from the intrinsic state $\vert \beta,\gamma=0;N\rangle$ of 
Eq.~(\ref{cond}),
\ba
\vert\, \beta; N, L,M\rangle &\propto&
\left [\Gamma_{N}^{(L)}(\beta)\right ]^{-1/2} 
\hat{\cal{P}}_{LM}\vert \beta,\gamma=0; N\rangle
\nonumber\\
\Gamma_{N}^{(L)}(\beta) &=& 
\frac{1}{N!}\int_{0}^{1}dx 
\left [ 1 + \beta^2\,P_{2}(x)\right ]^N\,P_{L}(x)
\label{wfqpt1}
\ea
where $P_{L}(x)$ is a Legendre polynomial with $L$ even
and $\Gamma_{N}^{(L)}(\beta)$ is a normalization factor.
The $L$-projected states $\vert\, \beta; N, L,M\rangle$ 
interpolate between the 
$U(5)$ spherical ground state, $\vert s^N\rangle$, with $n_d=\tau=L=0$, 
at $\beta=0$, 
and the $SU(3)$ deformed ground band with $(\lambda,\mu)=(2N,0)$, 
at $\beta=\sqrt{2}$. 
The matrix element of the Hamiltonian 
in these states define an $L$-projected energy surface 
which can be evaluated in closed form~\cite{lev05}
\ba
E^{(N)}_{L}(\beta) &=& 
\epsilon\left[\, N - S^{(N)}_{1,L}\,\right ]
+ \frac{1}{2}\kappa\,\left [\, 
(\beta^2 - 2)^2\, S^{(N)}_{2,L}
+ 2 (\beta - \sqrt{2})^2\, \Sigma^{(N)}_{2,L} + 
\frac{3}{4}L(L+1) - 2N(2N+3)\, \right ] ~.
\quad
\label{egL}
\ea
Here 
$S^{(N)}_{1,L}$, $S^{(N)}_{2,L}$, $\Sigma^{(N)}_{2,L}$ are, 
respectively, the expectation values of 
$\hat{n}_s=s^{\dagger}s$, $(s^{\dagger})^2s^2$, 
$\hat{n}_s\,\hat{n}_d$ in 
$\vert \beta;N,L,M\rangle$ and 
are given by
$S^{(N)}_{1,L} = \Gamma^{(L)}_{N-1}(\beta)/\Gamma^{(L)}_{N}(\beta)$, 
with $S^{(N)}_{2,L} = S^{(N)}_{1,L}S^{(N-1)}_{1,L}$ and
$\Sigma^{(N)}_{2,L} =
(N-1)S^{(N)}_{1,L} - S^{(N)}_{2,L}$. 
As shown in Fig.~(2b), 
the $L=0$ projected energy surface, $E^{(N)}_{L=0}(\beta)$, 
no-longer exhibits the double 
minima structure observed in the (unprojected) intrinsic energy 
surface. Instead, there is a minimum at $\beta>0$, 
a maximum at $\beta=0$, and a saddle point at $\beta<0$. 
The $L=2$ projected energy surface, 
$E^{(N)}_{L=2}(\beta)$, 
has a minimum at a larger value of $\beta>0$, 
and a flat shoulder near $\beta=0$. 
The different behaviour of 
$E^{(N)}_{L=2}(\beta)$ and $E^{(N)}_{L=0}(\beta)$ 
can be attributed to the fact that 
the $L=2$ state is 
well above the barrier 
and hence experiences essentially a flat-bottomed potential.
In contrast, 
the two minima in the intrinsic energy surface support two 
coexisting spherical and deformed $L=0$ states which 
are subject to considerable mixing. 
This mixing 
can be studied by means of a 
$2\times 2$ potential energy matrix, $K_{ij}$, calculated in the following 
orthonormal $L=0$ states 
\ba
\vert \Psi_1\rangle &=& \vert s^N\rangle 
\;\;\; , \;\;\;
\vert \Psi_2\rangle = 
(1-r_{12}^2)^{-1/2}\,
\Bigl (\,\vert \beta;N,L=0\rangle - r_{12}\,\vert s^N\rangle\, \Bigr )~,
\nonumber\\
K_{ij} &=& \langle \Psi_{i}\vert\, H_{cri}\,\vert \Psi_{j} \rangle 
\;\;\; , \;\;\;
r_{12} = \langle s^N\vert \beta;N,L=0\rangle = 
[N!\,\Gamma^{(L=0)}_{N}(\beta)]^{-1/2} ~.
\label{basis}
\ea
The derived eigenvalues of the matrix 
serve as eigenpotentials, $E^{(\pm)}_{L=0}(\beta)$, and the corresponding 
eigenvectors, $\vert \Phi^{(\pm)}_{L=0}\rangle$, 
are identified with the ground- and first-excited $L=0$ states. 
The deformed states $\vert \beta;N,L,M\rangle$ of Eq.~(\ref{wfqpt1}) 
with $L>0$ are identified 
with excited members of the ground-band with energies given by the 
L-projected energy surface $E_{L}^{(N)}(\beta)$, Eq.~(12).
As shown in Fig.~(2b), the lowest eigenpotential $E^{(-)}_{L=0}(\beta)$ 
has a global minimum at a certain $\beta>0$. Using this value in the
$L$-projected states leads, as seen in Table~2, to faithful 
estimates to the exact finite-N calculations of the critical
IBM Hamiltonian (notably for yrast states), which
in-turn capture the essential features of the $X(5)$
critical-point structure~\cite{iac01} relevant to $^{152}$Sm.

In a general first-order phase transition with an arbitrary barrier, 
the intrinsic energy energy surface of Eq.~(\ref{eint}) satisfies 
$a, b>0$ and $b^2=4ac$ at the critical-point. For $\gamma=0$
it can be transcribed in the form~\cite{lev05b}
\ba
E_{cri}(\beta) &=&
E_0 + c\,N(N-1)f(\beta)
\nonumber\\
f(\beta) &=& \beta^2\,(1+\beta^2)^{-2}\,(\beta - \beta_0)^2 ~.
\label{ecri1st}
\ea
As shown in Fig.~3, $E_{cri}(\beta)$ exhibits degenerate spherical and 
deformed minima, at $\beta=0$ and $\beta=\beta_0 =\frac{2a}{b} >0$. 
The value of $\beta_0$ determines the position $(\beta=\beta_{+})$ 
and height $(h)$ of the barrier separating the two minima in a 
manner given in the caption. 
To construct a critical Hamiltonian with such an energy surface, 
it is convenient to resolve it into intrinsic and 
collective parts~\cite{lev87},
\ba
H_{cri} = H_{int} + H_c ~.
\label{resol}
\ea
\begin{figure}[pt]  
\centerline{
\rotatebox{270}{\includegraphics[height=90mm]{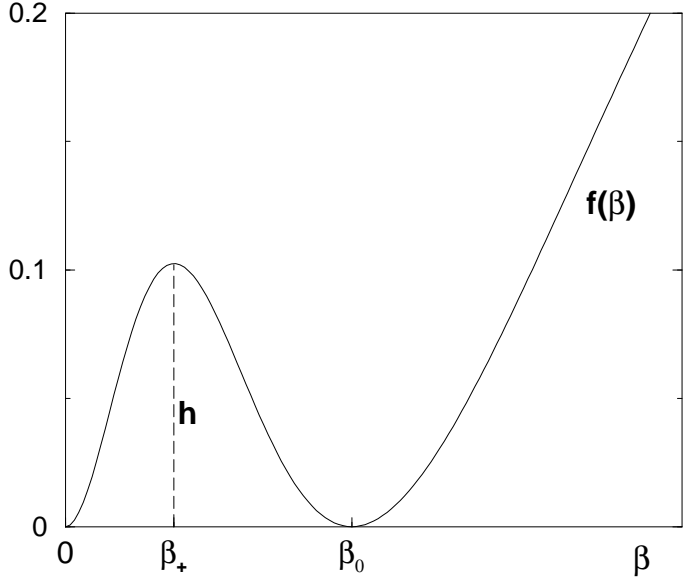}}}
\caption{The IBM energy surface, Eq.~(\ref{ecri1st}), 
at the critical-point of a 
first-order phase transition. 
The position and height of the barrier are 
$\beta_{+} = \frac{-1 + \sqrt{1+\beta_{0}^2}}{\beta_0}$ and  
$h = f(\beta_{+}) = 
\frac{1}{4}\,\left ( -1 + \sqrt{1+\beta_{0}^2}\;\right )^2$ respectively.
\label{f1}}
\end{figure}
The intrinsic part 
($H_{int}$) 
is defined to have the equilibrium 
condensate $\vert \,\beta=\beta_0,\gamma=0 ; N\rangle$, Eq.~(\ref{cond}), 
as an exact zero-energy eigenstate and to have an energy surface as in 
Eq.~(14) (with $E_0=0$). 
$H_{int}$ has the form
\ba
H_{int} &=& h_{2}\,
\left [\beta_{0}\,s^{\dagger}d^{\dagger} + 
\sqrt{7/2}\,\left( d^{\dagger} d^{\dagger}\right )^{(2)}\right ]\cdot
\left [\beta_{0}\,\tilde{d}s + 
\sqrt{7/2}\,( \tilde{d}\, \tilde{d})^{(2)}\right ] ~,
\label{hint}
\ea
and by construction has the $L$-projected states 
$\vert \beta=\beta_0;N,L\rangle$ of Eq.~(\ref{wfqpt1}) 
as solvable deformed eigenstates with energy $E=0$. 
It has also solvable spherical eigenstates:  
$\vert N,n_d=\tau=L=0 \rangle\equiv \vert s^N\rangle$ and 
$\vert N,n_d=\tau=3,L=3 \rangle$ with energy $E=0$ 
and $E = 3 h_2\left [\beta_{0}^2 (N-3) + 5 \right ]$ respectively.
For large $N$ the spectrum of $H_{int}$ is harmonic, involving 
quadrupole vibrations about the spherical minimum with frequency 
$\epsilon$, and both $\beta$ and $\gamma$ vibrations about the deformed 
minimum with frequencies 
$\epsilon=\epsilon_{\beta}=h_{2}\,\beta_{0}^2 N \gg 
\epsilon_{\gamma} = 9(1+\beta_{0}^2)^{-1}\,\epsilon_{\beta}$, 
where the last inequality holds in 
the acceptable range $0\leq\beta_0\leq 1.4$. 
All these features are present in the exact spectrum of $H_{int}$ shown 
in Fig.~4, which 
displays a zero-energy deformed ($K=0$) ground band, degenerate with a 
spherical $(n_d=0)$ ground state. The remaining states are either 
predominantly spherical, or deformed states arranged in several excited 
$K=0$ bands below the $\gamma$ band. The coexistence of spherical and 
deformed states is evident in the right portion of Fig.~4, which shows 
the $n_d$ decomposition of wave functions of selected eigenstates of 
$H_{int}$. The ``deformed'' states show a broad $n_d$ distribution typical 
of a deformed rotor structure. The ``spherical'' states show the 
characteristic dominance of single $n_d$ components that one would expect 
for a spherical vibrator. 

The collective part ($H_c$) of the full critical Hamiltonian, Eq.~(15), 
is composed of kinetic terms which do not affect the shape of the 
intrinsic energy surface. It 
can be transcribed in the form~\cite{lev87}
\ba
H_{c} &=& c_3 \left [\, \hat{C}_{O(3)} - 6\hat{n}_d \,\right ]
+ c_5 \left [\, \hat{C}_{O(5)} - 4\hat{n}_d \,\right ] 
+ c_6 \left [\, \hat{C}_{\overline{O(6)}} - 5\hat{N}\,\right ] +E_0~,
\label{hcol}
\ea
where $\hat{N}=\hat{n}_d+\hat{n}_s$. 
Here $\hat{C}_{G}$ denotes the quadratic Casimir operator of the 
group G as defined in \cite{lev87}. 
Table~3 shows the 
effect of different rotational terms in $H_c$. 
For the high-barrier case considered here, ($\beta_0=1.3$, $h=0.1$), 
the calculated spectrum resembles a rigid-rotor $(E\sim a_{N}L(L+1)$) for 
the $c_3$-term, a rotor with centrifugal stretching 
$(E\sim a_{N}L(L+1) - b_{N}[L(L+1)]^2)$ for the $c_5$-term, and a 
X(5)-like spectrum for the $c_6$-term. In all cases the B(E2) values are 
close to the rigid-rotor Alaga values. This behaviour is different from 
that encountered when the barrier is low, {\it e.g.}, for the 
$U(5)$-$SU(3)$ case discussed above, corresponding to 
$\beta_0=1/2\sqrt{2}$ and $h\approx 10^{-3}$, 
where both the spectrum and E2 transitions are similar to the X(5) 
predictions. 
\begin{table}[h]
\centering
\caption{
Excitation energies and B(E2) values (in units as in Table 2)
for selected terms in the critical Hamiltonian, 
Eqs.~(\ref{resol}),(\ref{hint}),(\ref{hcol}).
The exact values are for $N=10$, $\beta_0=1.3$
and $c_L/h_2$ coefficients indicated in the Table. 
The entries in square brackets [...] are estimates based on the 
$L$-projected states, Eq.~(\ref{wfqpt1}),  
with $\beta=1.327,\,1.318,1.294$, determined by the global minimum of 
the respective lowest eigenvalue of the potential matrix, 
Eqs.~(\ref{Kij}),(\ref{eneL}). 
The rigid-rotor and X(5)~{\protect\cite{iac01}} predictions are 
shown for comparison.
\normalsize}
\vskip 10pt
\begin{tabular}{lccccc}
\hline\hline\noalign{\smallskip}
                & $c_3/h_2=0.05$ & $c_5/h_2 = 0.1$ & $c_6/h_2=0.05$ & 
rotor & X(5) \\
\noalign{\smallskip}\hline\noalign{\smallskip}
$E(4^{+}_{1})$  & 
           3.32  [3.32]  & 3.28 [3.28]   & 2.81 [2.87]    & 3.33  & 2.91 \\
$E(6^{+}_{1})$  & 
           6.98  [6.97]  & 6.74 [6.76]   & 5.43 [5.63]    & 7.00  & 5.45 \\
$E(8^{+}_{1})$  & 
          11.95 [11.95]  & 11.23 [11.29] & 8.66 [9.04]    & 12.00 & 8.51 \\
$E(10^{+}_{1})$ & 
          18.26 [18.26]  & 16.58 [16.69] & 12.23 [12.83]  & 18.33 & 12.07 \\
$E(0^{+}_{2})$  & 
           6.31  [6.30]  & 6.01 [5.93]   & 4.56 [5.03]    &       & 5.67 \\
\noalign{\smallskip}\hline\noalign{\smallskip}
$B(E2; \, 4^{+}_{1}\to 2^{+}_{1})$ &
           1.40 [1.40]   & 1.40 [1.40]  & 1.45 [1.44]  & 1.43  & 1.58 \\
$B(E2;\, 6^{+}_{1}\to 4^{+}_{1})$ & 
           1.48 [1.48]   & 1.48 [1.48]  & 1.53 [1.52]  & 1.57  & 1.98 \\
$B(E2;\, 8^{+}_{1}\to 6^{+}_{1})$ & 
           1.46 [1.45]   & 1.46 [1.45]   & 1.50 [1.50]  & 1.65  & 2.27 \\
$B(E2;\, 10^{+}_{1}\to 8^{+}_{1})$  & 
           1.37 [1.37]   & 1.38 [1.37]  & 1.41 [1.55]  & 1.69  & 2.61 \\
$B(E2;\, 0^{+}_{2}\to 2^{+}_{1})$ & 
           0.005 [0.007] & 0.006 [0.007] & 0.19 [0.16] &       & 0.63 \\
\noalign{\smallskip}\hline\hline
\end{tabular}
\end{table}
\begin{figure}[pt]  
\centering
\rotatebox{270}{\includegraphics[width=0.35\linewidth]{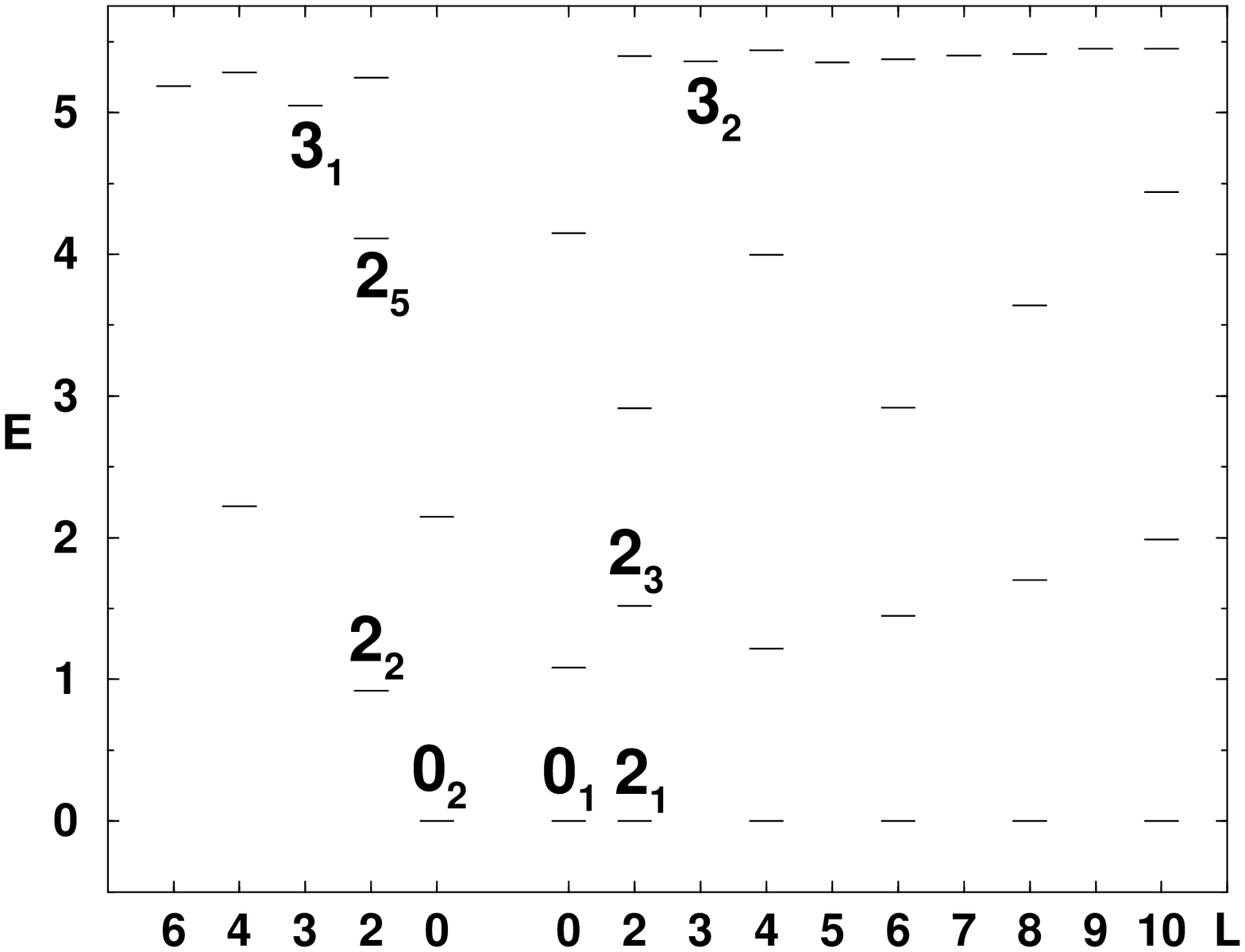}}
\rotatebox{270}{\includegraphics[width=0.39\linewidth]{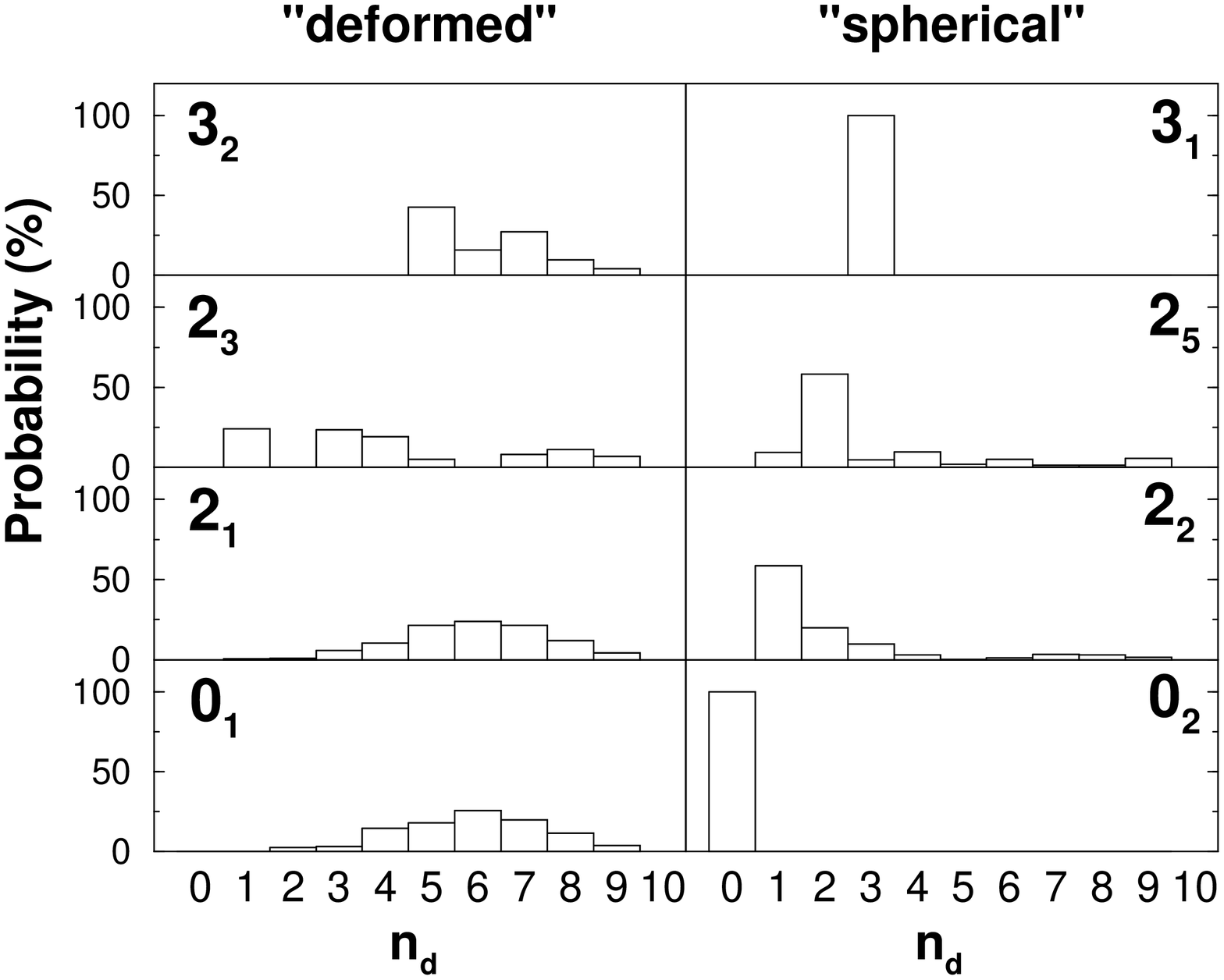}}
\caption{Left potion: spectrum of $H_{int}$, Eq.~(\ref{hint}), 
with $h_2=0.1$, $\beta_0 =1.3$ and $N=10$. 
Right~portion: the number of $d$ bosons ($n_d$) 
probability distribution for selected eigenstates of $H_{int}$.
\label{f4}}
\end{figure}

Considerable insight of the underlying 
structure at the critical-point is gained by examining the 
$2\times 2$ potential energy matrix, $K_{ij}$, of Eq.~(\ref{basis}),  
which reads
\ba
K_{11} &=& E_0\;\; , \;\; 
K_{12} = -c_6\,\beta^2 N(N-1)(1-r_{12}^2)^{-1/2}r_{12} ~,
\nonumber\\
K_{22} &=& E_0 + (1-r_{12}^2)^{-1}
\left [\,\tilde{E}^{(N)}_{L=0}(\beta) 
+ 2c_6\,\beta^2N(N-1)r_{12}^2\, \right ] ~.
\label{Kij}
\ea
Apart from a constant shift, $\tilde{E}_{L}^{(N)}(\beta)$ is 
the $L$-projected energy surface of $H_{cri}$~(\ref{resol}), and 
is given by
\ba
\tilde{E}_{L}^{(N)}(\beta) = E_{L}^{(N)}(\beta) - E_0 &=&   
h_2\,(\beta-\beta_0)^2\Sigma_{2,L}^{(N)} + 
c_3\left [ L(L+1) - 6D_{1,L}^{(N)}\right ]
+ c_5\left [ -\beta^4\,S_{2,L}^{(N)} + D_{2,L}^{(N)}\right ]
\nonumber\\
&&
+\, c_6\left [ -(1+\beta^2)^2\,S_{2,L}^{(N)} +N(N-1)\right ] ~.
\label{eneL}
\ea
Here 
$\Sigma_{2,L}^{(N)}$, $D_{1,L}^{(N)}$, $D_{2,L}^{(N)}$ and  
$S_{2,L}^{(N)}$ denote the expectation values of $\hat{n}_s\hat{n}_d$, 
$\hat{n}_d$, $\hat{n}_d(\hat{n}_d-1)$ and 
$\hat{n}_s(\hat{n}_s-1)$ respectively in 
$\vert \beta;N,L,M\rangle$ (\ref{wfqpt1}).
As seen in Table~3, by determining the value of $\beta$ in the 
$L$-projected states from the 
global minimum of the lowest eigenvalue of the potential matrix 
(\ref{Kij}), one obtains 
accurate finite-N 
estimates to the energies and E2 rates at the critical-point. 

Part of the work reported was done in collaboration with J.N. Ginocchio 
(LANL). This work was supported by the Israel Science Foundation.

\end{document}